# Misestimation of temperature when applying Maxwellian distributions to space plasmas described by kappa distributions

Georgios Nicolaou[1], George Livadiotis[2]

1. Swedish Institute of Space Physics, Kiruna, Sweden
2. Southwest Research Institute, San Antonio, USA

**Abstract**

This paper presents the misestimation of temperature when observations from a kappa distributed plasma are analyzed as a Maxwellian. One common method to calculate the space plasma parameters is by fitting the observed distributions using known analytical forms. More often, the distribution function is included in a forward model of the instrument's response, which is used to reproduce the observed energy spectrograms for a given set of plasma parameters. In both cases, the modeled plasma distribution fits the measurements to estimate the plasma parameters. The distribution function is often considered to be Maxwellian even though in many cases the plasma is better described by a kappa distribution. In this work we show that if the plasma is described by a kappa distribution, the derived temperature assuming Maxwell distribution can be significantly off. More specifically, we derive the plasma temperature by fitting a Maxwell distribution to pseudo-data produced by a kappa distribution, and then examine the difference of the derived temperature as a function of the kappa index. We further consider the concept of using a forward model of a typical plasma instrument to fit its observations. We find that the relative error of the derived temperature is highly depended on the kappa index and occasionally on the instrument's field of view and response.

## 1. Introduction

Several instruments observe the plasma in a wide range of space environments. Typical plasma instruments obtain the energy spectrograms of the plasma, i.e.,, they measure the rate of particles ($C/s$) entering the detection system as a function of energy ($E$). It is often useful and desirable to derive the fluid properties of the observed plasma for further scientific analysis. Space plasma fluid parameters can be quantified from instrument's observations using several techniques. Each technique is chosen based on the plasma environment and the characteristics of the instrument, such as, the Field Of View (FOV) and the angular resolution. The specific analytical form of the velocity or the energy distribution is required in order to apply some of the plasma parameter derivation techniques.

The two types of distribution functions that are commonly used to describe space plasma populations are the kappa and Maxwell distributions of velocities (Livadiotis 2015a). Maxwell distributions have been observed in space plasmas (e.g., Hammond et al. 1996). Although Maxwell distributions are often used in space science because they fit well the "core" of the observed distributions, numerous studies have shown that kappa distributions (or combinations thereof) are frequently observed (Shizgal 2007; Livadiotis and McComas 2009, 2013a; Pierrard and Lazar 2010; Livadiotis 2015a; and references therein). Kappa distributions are characterized by their high energy tail which is not always clearly observed because of the low plasma intensities in the high energy range. Nevertheless, many analyses used successfully kappa distributions to describe observations in several plasma environments, where the Maxwell distribution clearly fails to describe the high energy tails of the observed distributions: solar wind (e.g., Maksimovic et al. 1997; 2005; Pierrard et al. 1999; Chotoo et al. 2000; Mann et al. 2002; Marsch 2006; Zouganelis et al. 2008; Štverák et al. 2009; Livadiotis and McComas 2013b; Yoon 2014; Heerikhuisen et al 2015; Pierrard and Pieters 2015), planetary magnetospheres (e.g., Christon 1987; Collier and Hamilton 1995; Jurac et al. 2002; Pisarenko et al. 2002; Kletzing et al. 2003; Mauk et al. 2004; Schippers et al. 2008; Dialynas et al. 2009; Ogasawara et al. 2012; Carbary et al. 2014; Qureshi et al. 2015; Stepanova and Antonova 2015), the outer heliosphere and the inner heliosheath (e.g., Decker and Krimigis 2003; Decker et al. 2005; Heerikhuisen et al 2008; 2010; Zank et al. 2010; Livadiotis et al. 2011; 2012; 2013; Livadiotis and McComas 2011a; 2012), and other various plasma-related analyses (e.g., Milovanov and Zelenyi 2000; Saito et al. 2000; Yoon et al. 2006; Raadu and Shafiq 2007; Livadiotis 2009; Tribeche et al. 2009; Hellberg et al. 2009; Livadiotis and McComas 2009; Baluku et al. 2010; Livadiotis and McComas 2010a; 2010b; Le Roux et al. 2010; Eslami et al. 2011; Livadiotis and McComas 2011b; Kourakis et al. 2012; Livadiotis and McComas 2013c; Livadiotis 2014; 2015b; 2015c; 2016a; 2016b; Randol and Christian 2014; Varotsos et al. 2014; Liu et al. 2015; Fisk and Gloeckler 2015; Viñas et al. 2015). In addition, several physical mechanisms that can be applied in different plasma environments, successfully explain the observed deviation from Maxwell distribution. For example, it has been shown that kappa distribution can be generated from velocity-space diffusion processes (Hasegawa et al. 1985), pick-up ions (Livadiotis & McComas 2011a), Langmuir turbulence (Yoon 2012), etc. Besides their empirical successful usage and their extraction from different mechanisms, kappa distributions are naturally exported (Milovanov and Zelenyi 2000; Leubner 2000; Livadiotis and McComas 2009; Livadiotis 2015a) from the foundations of Tsallis non-extensive Statistical mechanics (Tsallis 1988; 2009; Tsallis et al. 1999).

The benefit of the kappa distribution is that it can successfully describe plasmas that are out of thermal equilibrium while Maxwell distribution, as a limiting case of kappa distribution, describes only plasmas that are in thermal equilibrium. In cases where the plasma parameters are derived from a method that uses a specific type of the distribution function, the wrong selection of the distribution may lead to significant misestimation of the plasma parameters; especially, of the plasma temperature.

In this paper we examine two methods that are widely used to derive the plasma properties from observations, that is (i) direct fitting of the distribution function, where an analytical expression of the distribution function is used (e.g., Paschmann and Daly 1998; Wilson et al. 2012a); and (ii) forward modeling, which is used to fit the plasma instrument's observations with an expression that involves, among others, the distribution function and the instrument's response function (e.g., Richardson 1987; 2002; Wilson et al. 2008; Elrod et al. 2012; Wilson et al. 2012b; 2013; Livi et al. 2014; Nicolaou et al. 2014; 2015a; 2015b; Elliott et al. 2016). In both methods we examine here, the form of the distribution function has to be defined before its application to the data. For the purposes of this study, we simulate the plasma measurements for fixed plasma parameters. Then each method is applied to derive the plasma temperature which is then compared to the plasma temperature that is set in the simulations. The derived temperature misestimation is quantified as a function of several parameters, including some characteristics of the instrument that is used for the plasma measurements. This paper is organized as follows: In section 2 we describe the two methods we investigate: (i) direct fitting and (ii) forward modeling. In section 3 we present our results using simulated measurements, while in section 4 we summarize and discuss the conclusions.

## 2. Methods of plasma parameters derivation

In general, the choice of the method to be used is based usually on the available measurements and/or the instrument's characteristics. In this section we present the two methods we investigate.

### *2.1 Fitting of the distribution function*

There are cases, depending on the plasma environment and the instrument's characteristics, where the energy spectrograms $C(E)$ can be directly converted to the energy distribution function $f(E)$ or the velocity distribution function $f(\vec{u})$ of the observed plasma (e.g., Paschmann and Daly 1998; Wilson et al. 2012a). In those cases we can get the plasma parameters by taking the statistical moments of the distribution function (by numerical integration) or by fitting the observed distribution. For example, the first-order moment (the mean) gives the bulk velocity, the second-order moment around the mean, leads to the temperature, etc. On the other hand, these parameters, the bulk velocity and the temperature, can be derived by fitting a distribution function to the data. In order to fit the observations, we need to specify the analytical expression of the distribution as a function of the plasma parameters. As mentioned previously, the most common energy distribution functions that are used to describe space plasma are the kappa distribution

$$f_\kappa(E,\omega) = \frac{2}{\sqrt{\pi}} \cdot n (k_B T)^{-\frac{3}{2}} A_\kappa \cdot \left[ 1 + \frac{E + E_0 - 2\sqrt{E E_0} \cos \omega}{(\kappa - \frac{3}{2}) k_B T} \right]^{-\kappa - 1}, \quad (1a)$$

with:

$$A_\kappa = (\kappa - \tfrac{3}{2})^{-\frac{3}{2}} \frac{\Gamma(\kappa + 1)}{\Gamma(\kappa - \tfrac{1}{2})}, \quad (1b)$$

and the Maxwell distribution

$$f_M(E,\omega) = \frac{2}{\sqrt{\pi}} \cdot n (k_B T)^{-\frac{3}{2}} \cdot \exp\left( -\frac{E + E_0 - 2\sqrt{E E_0} \cos \omega}{k_B T} \right), \quad (2)$$

where the normalization constants are derived from

$$\tfrac{1}{n} \int_0^\infty f(\varepsilon) \varepsilon^{\frac{1}{2}} d\varepsilon = 1, \quad (3)$$

with

$$\varepsilon \equiv \tfrac{1}{2}m(\vec{u}-\vec{u}_b)^2 = E + E_0 - 2\sqrt{EE_0}\cos\omega ,\qquad(4)$$

denoting the energy of the particle in the reference frame that moves with the bulk speed. In the above, $E_0 = \tfrac{1}{2}m\vec{u}_b^{\,2}$, $T$, and $n$ are the plasma bulk energy, temperature, and density respectively; $k_B$ is the Boltzman constant; the variable $\omega$ reads the angle between a plasma particle velocity and the flow vector. The kappa index $\kappa$ is (i) characteristic of the state of plasma indicating a measure of its "thermodynamic distance" from thermal equilibrium (Livadiotis and McComas 2010a), and (ii) interwoven with the statistical correlation between the energy of the particles (Livadiotis 2015a; 2015c). The Maxwell distribution describes the special case when the kappa value of the distribution is infinity (thermal equilibrium); the other extreme state, where kappa approaches its minimum value ($\kappa \to 3/2$), describes the furthest state from thermal equilibrium called "the anti-equilibrium" that is the specific state of the system where the kappa index tends to its lowest value ($\kappa \to 3/2$) (e.g., Livadiotis and McComas 2010a; 2013a; Livadiotis 2015a).

The plasma parameters (e.g., the density, temperature, and kappa index) can be determined from the fitting of analytical distribution functions to observed datasets.

## *2.2 Fitting of the forward modeling*

There are cases where the distribution function of the plasma is not directly derived from the instrument's spectrograms. In those cases, a forward modeling of the instrument's response is used to fit the observed energy spectrograms for specific set of the plasma parameters (e.g., Richardson 1987; 2002; Wilson et al. 2008; Elrod et al. 2012; Wilson et al. 2012b; 2013; Livi et al. 2014; Nicolaou et al. 2014; 2015a; 2015b; Elliott et al. 2016). Such models predict the observed counts that will be observed, for specific plasma conditions and taking into account the instrument's characteristics and response. Typical plasma instruments give the number of counts for discrete energy steps. The general expression for the counts measured at each energy step $C(E)$ is

$$C(E) = \frac{2\Delta t}{m^2} \cdot \int_{E-\tfrac{1}{2}\Delta E}^{E+\tfrac{1}{2}\Delta E} \int R(E';\Omega) E' f(E';\Omega) d\Omega dE' ,\qquad(5)$$

where the solid angle $\Omega$ comprises both the polar and azimuth angles ($\vartheta,\varphi$), integrated over the whole FOV of the instrument; the channel energy range $\Delta E$ is defined by the geometric characteristics of the instrument. The factor $R$ is characteristic of the instrument, a function of the instrument's aperture, efficiency and response that can be determined through the instrument's calibration. The time resolution of the instrument is denoted by $\Delta t$. For an electrostatic analyzer type of instruments the energy resolution $\Delta E/E$ (or, $\Delta \ln E$) is constant and in cases where it is small enough such we can rewrite Eq.(5) as

$$C(E) \cong \frac{2\Delta t \Delta \ln E}{m^2} \cdot \int R(E;\Omega) E^2 f(E;\Omega) d\Omega .\qquad(6)$$

For both methods described above, the choice of the distribution function affects the determination of the plasma parameters and especially the plasma temperature.

For the purposes of this study, we model observations of plasma that is described by a kappa distribution and fixed plasma parameters. We then quantified the miss-estimation of the plasma temperature when the plasma parameters are derived using the methods described above but with a Maxwellian distribution instead of a kappa distribution.

**3. Results**

Each method is examined as it is applied on simulated measurements of plasma with fixed parameters. In this section we present for each method, the misestimation of the plasma temperature as a function of several parameters.

*3.1. Temperature error when direct fitting is applied*

We firstly consider the case where the observed plasma is described by kappa distribution and the parameters are calculated by direct fitting of an analytical expression to the observed distribution. We would like to show how someone can misestimate the plasma temperature by fitting Maxwell distribution to the data instead of kappa distribution. A proper fitting of distribution to the data is achieved using the chi-square minimization technique. If the plasma is described by a kappa distribution, then only the core of the distribution will be well-fitted by the Maxwellian distribution, leading to a misestimation of the plasma temperature (Livadiotis and McComas 2009; 2013a). To show that we rewrite Eq.(1) as

$$f_\kappa(\varepsilon) = C_\kappa \left[1 + \frac{\varepsilon}{(\kappa - \frac{3}{2})k_B T}\right]^{-\kappa-1}, \qquad (7)$$

where the normalization constant $C_\kappa = \frac{2}{\sqrt{\pi}} \cdot n(k_B T)^{-\frac{3}{2}} A_\kappa$. Expanding its logarithm for small particle energies in the co-moving reference frame, $\varepsilon \ll (\kappa - \frac{3}{2})k_B T$,

$$\ln f_\kappa(\varepsilon) = \ln C_\kappa - (\kappa+1)\ln\left[1 + \frac{\varepsilon}{(\kappa - \frac{3}{2})k_B T}\right] \cong \ln C_\kappa - \left(\frac{\kappa+1}{\kappa - \frac{3}{2}}\right) \cdot \frac{\varepsilon}{k_B T}, \qquad (8)$$

which leads to a Maxwell distribution,

$$f_\kappa(\varepsilon) \cong C_\kappa e^{-\frac{\varepsilon}{k_B T_M}}, \qquad (9)$$

with temperature

$$T_M = \left(\frac{\kappa - \frac{3}{2}}{\kappa + 1}\right) \cdot T. \qquad (10)$$

We make it clear that the condition $\varepsilon \ll (\kappa - \frac{3}{2})k_B T$ means that, for any given kappa index and temperature, that is, for any value of the product $(\kappa - \frac{3}{2})k_B T$, there is always a range of energy that the approximation is satisfactory, and that is $\varepsilon \ll (\kappa - \frac{3}{2})k_B T$. Given the Taylor expansion

$$\ln\left[1+\frac{\varepsilon}{\left(\kappa-\tfrac{3}{2}\right)k_BT}\right] \cong \frac{1}{\kappa-\tfrac{3}{2}}\cdot\frac{\varepsilon}{k_BT} - \frac{1}{2(\kappa-\tfrac{3}{2})^2}\cdot\left(\frac{\varepsilon}{k_BT}\right)^2 + O\left[\frac{\varepsilon}{(\kappa-\tfrac{3}{2})k_BT}\right]^3,$$

then, the relative error of keeping just the first term in (8) is:

$$\text{Relative Error} \equiv 2^{\text{nd}} \text{ term} / 1^{\text{st}} \text{ term} \sim \frac{\varepsilon}{2(\kappa-\tfrac{3}{2})k_BT}$$

For example, let $\kappa\sim 2$, $k_BT \sim 10\,\text{eV}$ and $E_0 \equiv \tfrac{1}{2}m\vec{u}_b^{\,2} \sim 1\,\text{keV}$ which corresponds to $u_b \sim 440\,\text{km/s}$ in the case we study the distribution function of protons; hence, relative error $\sim 100\,\varepsilon/E_0$. For having relative error $<1$, we need $\varepsilon < E_0/100$ or $|\vec{u}-\vec{u}_b| < u_b/10$; e.g., for $u_b \sim 440\,\text{km/s}$, we have $400\,\text{km/s} < u < 480\,\text{km/s}$, where the core is well approximated by a Maxwellian velocity distribution with the temperature given by (10). In order to demonstrate that, in **Figure 1**, we show two cases of kappa distributions fitted by Maxwellian distributions and the relative error given above, as a function of energy. It is shown that there is an energy range for which the Maxwellian approach of the distribution's "core" is valid, even for small kappa ($\kappa \to 2$) indices and for relatively cold plasma ($T \to 10\,\text{eV}$), which are respectively the lowest kappa and temperature we examine in this study. As expected, for larger kappa the relative error is quite small within a wider range of energies.

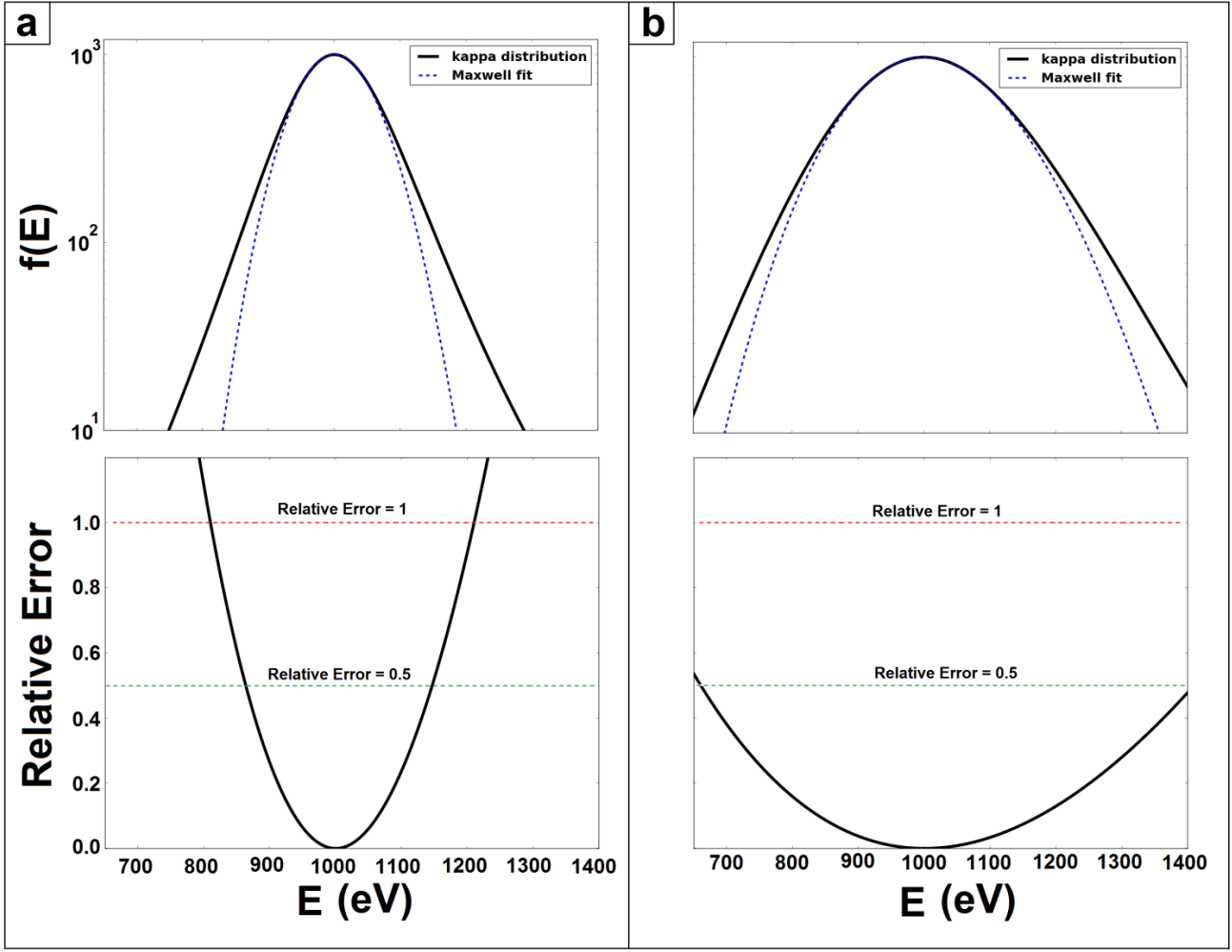

**Figure 1 a)** (top) Kappa distribution for $\kappa=2$, $E_0=1$keV, $T=10$eV and a Maxwellian fit to the distribution's "core" and (below) the relative error of keeping just the first term of the expansion in (8), as a function of energy. The green and red dashed lines indicate where the relative error is 0.5 and 1 respectively. **b)** The same plots for $\kappa=5$, $E_0=1$keV, $T=10$eV. In this case, the distribution's "core" is wider and the relative error is significantly smaller within a wide range of energies. In both cases though, there is a reasonable range where the distribution's core can be approached by a Maxwellian distribution with temperature given by (10).

Assuming then, that the plasma distributions are fitted with a Maxwellian expression instead of a kappa distribution and the fitting algorithm is forced to find the best fit near the distribution's maximum where the statistical error is small (assuming Poisson statistics for the measurements), it is shown that the misestimation of the temperature is given by Eq.(10). It is directly noticeable that the derived temperature approaches the actual plasma temperature when this is in thermal equilibrium, $\kappa\rightarrow\infty$, but it is significantly off when the plasma approaches the furthest state from thermal equilibrium, the anti-equilibrium, where, $\kappa$, approaches its lowest value of 3/2 (Livadiotis 2015a). In order to demonstrate the temperature misestimation in practice, we run multiple fits of different kappa distributions using a Maxwell distribution model. We consider the distribution along the direction of the bulk speed ($\omega=0°$) and we fit the highest counts which are near the distribution's maximum (the "core" of the distribution), since those data-points are statistically more significant. In the upper panel of **Figure 2** we show few examples of a Maxwell fit to kappa distributions. For all the presented examples the bulk energy of the distributions is $E_0=1$ keV while the density is adjusted such us all of the compared distributions have the same maximum value. We then examine the derived from fitting Maxwellian temperature ($T_{M,F}$) as a function of the kappa index and for several plasma temperatures ($T$). The plots of the derived $T_{M,F}$ as a function of kappa index for each plasma temperature are shown in the lower panel of **Figure 2**. On each plot we draw also the curve of Eq. (10) which gives the theoretically misestimated temperature ($T_M$). In Table 1 the reader can find the derived $T_{M,F}$ values as a function of kappa for each $T$, while a straight comparison is made with the theoretically expected value ($T_M$). We see that the derived from fitting $T_{M,F}$ follows very well the theoretically derived curve, especially at higher kappa values. Note that, we do not examine cases for $\kappa<2$, because as we approach the limit of $\kappa\rightarrow 3/2$, the distribution's core shifts to very low energies (e.g., see Fig.1 in Livadiotis 2014) and practically there is no reasonable energy range to fit a Maxwellian (in such a case we would have $T_M\rightarrow 0$).

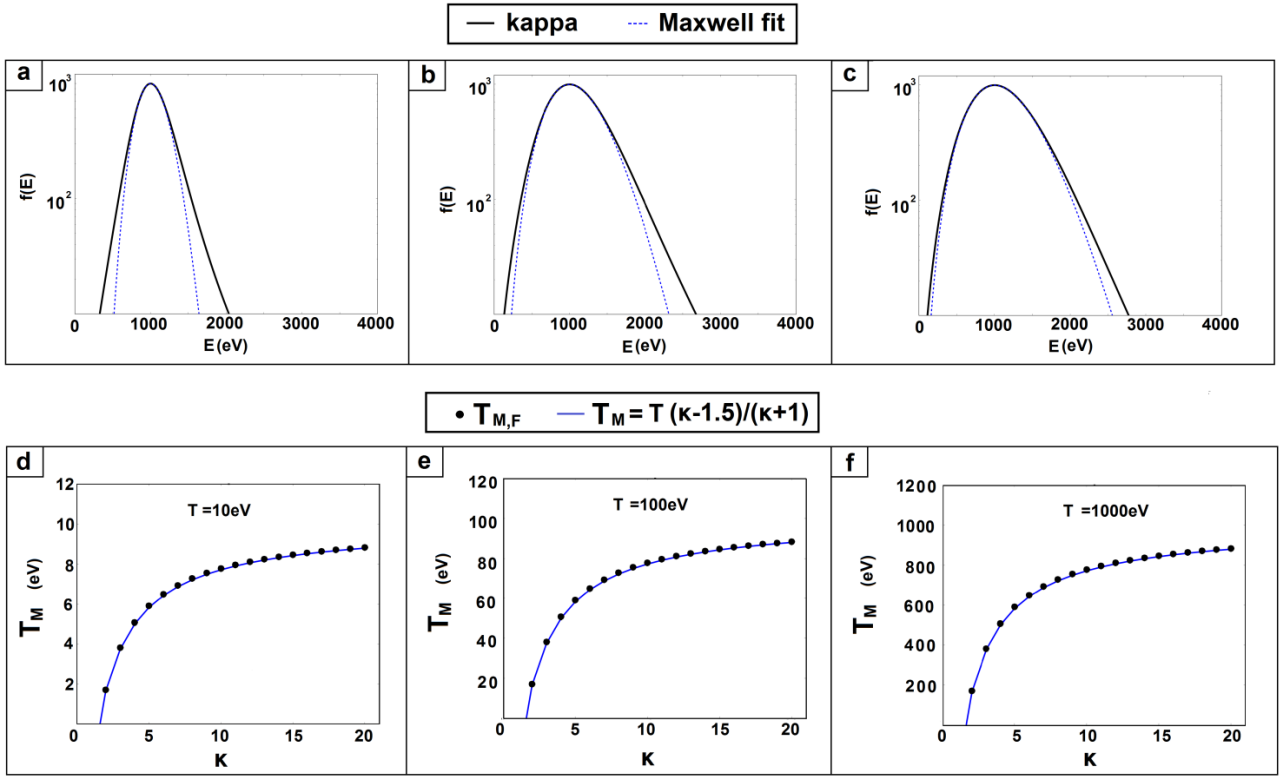

**Figure 2** The Maxwell distribution (blue dashed line) is fitted to kappa distribution (black solid line) for kappa index (a) $\kappa=2$, (b) $\kappa=5$ and (c) $\kappa=10$. For the presented examples we consider the distribution along the bulk velocity direction ($\omega=0°$). The bulk energy is $E_0=1$ keV and the temperature is $T=100$ eV/$k_B$, while the density is adjusted, such us, all the distributions have the same maximum value (arbitrarily chosen to be 1000). The mistaken temperature as derived by Maxwell fit to the kappa distribution is shown as a function of the kappa index for plasma temperature (d) $T=10$ eV/$k_B$, (e) $T=100$ eV/$k_B$ and (f) $T=1000$ eV/$k_B$ respectively. The solid blue line depicts the analytical model of $T_M$ given by Eq.(10), while the dots show the temperature derived by fitting a Maxwellian ($T_{M,F}$). We demonstrate that the formula of Livadiotis and McComas (2009; 2013) provides correctly the misestimation on temperature.

| | Plasma Temperature $T$=10eV | | | | Plasma Temperature $T$=100eV | | | | Plasma Temperature $T$=1000eV | | |
|---|---|---|---|---|---|---|---|---|---|---|---|
| $\kappa$ | $T_{M,F}$ (eV) | $T_M$ (eV) | $\frac{T_{M,F}-T_M}{T_M}$ (%) | $\kappa$ | $T_{M,F}$ (eV) | $T_M$ (eV) | $\frac{T_{M,F}-T_M}{T_M}$ (%) | $\kappa$ | $T_{M,F}$ (eV) | $T_M$ (eV) | $\frac{T_{M,F}-T_M}{T_M}$ (%) |
| 2 | 1.71 | 1.67 | 2.67 | 2 | 17.11 | 16.67 | 2.68 | 2 | 171.16 | 166.67 | 2.69 |
| 3 | 3.83 | 3.75 | 2.01 | 3 | 38.25 | 37.50 | 2.01 | 3 | 382.55 | 375.00 | 2.01 |
| 4 | 5.08 | 5.00 | 1.56 | 4 | 50.80 | 50.00 | 1.61 | 4 | 508.04 | 500.00 | 1.61 |
| 5 | 5.91 | 5.83 | 1.35 | 5 | 59.11 | 58.33 | 1.31 | 5 | 591.14 | 583.33 | 1.34 |
| 6 | 6.50 | 6.43 | 1.12 | 6 | 65.02 | 64.29 | 1.14 | 6 | 650.21 | 642.86 | 1.14 |
| 7 | 6.94 | 6.88 | 0.97 | 7 | 69.43 | 68.75 | 0.99 | 7 | 694.38 | 687.50 | 1.00 |
| 8  | 7.29 | 7.22 | 0.87 | 8  | 72.86 | 72.22 | 0.88 | 8  | 728.65 | 722.22 | 0.89 |
| 9  | 7.56 | 7.50 | 0.80 | 9  | 75.60 | 75.00 | 0.80 | 9  | 756.01 | 750.00 | 0.80 |
| 10 | 7.78 | 7.73 | 0.73 | 10 | 77.83 | 77.27 | 0.72 | 10 | 778.35 | 772.73 | 0.73 |
| 11 | 7.97 | 7.92 | 0.65 | 11 | 79.69 | 79.17 | 0.66 | 11 | 796.94 | 791.67 | 0.67 |
| 12 | 8.13 | 8.08 | 0.61 | 12 | 81.26 | 80.77 | 0.61 | 12 | 812.66 | 807.69 | 0.62 |
| 13 | 8.26 | 8.21 | 0.57 | 13 | 82.60 | 82.14 | 0.56 | 13 | 826.13 | 821.43 | 0.57 |
| 14 | 8.38 | 8.33 | 0.52 | 14 | 83.77 | 83.33 | 0.53 | 14 | 837.78 | 833.33 | 0.53 |
| 15 | 8.48 | 8.44 | 0.50 | 15 | 84.79 | 84.38 | 0.49 | 15 | 847.97 | 843.75 | 0.50 |
| 16 | 8.57 | 8.53 | 0.46 | 16 | 85.69 | 85.29 | 0.46 | 16 | 856.96 | 852.94 | 0.47 |
| 17 | 8.65 | 8.61 | 0.44 | 17 | 86.49 | 86.11 | 0.44 | 17 | 864.94 | 861.11 | 0.44 |
| 18 | 8.72 | 8.68 | 0.42 | 18 | 87.20 | 86.84 | 0.42 | 18 | 872.07 | 868.42 | 0.42 |
| 19 | 8.78 | 8.75 | 0.39 | 19 | 87.85 | 87.50 | 0.39 | 19 | 878.50 | 875.00 | 0.40 |
| 20 | 8.84 | 8.81 | 0.37 | 20 | 88.43 | 88.10 | 0.38 | 20 | 884.31 | 880.95 | 0.38 |

**Table 1** The derived temperature $T_{M,F}$ as a function of κ for the examples shown in **Figure 2**. For straight comparison with the theory we show the expected value ($T_M$) as calculated from equation (8) and the difference of the derived and expected value normalized to the expected one.

### *3.2 Temperature error when forward modeling is applied*

In many cases the distribution function *f(E)* cannot be directly constructed from the observations. In those cases we can develop a forward model of the instrument's response and reproduce the observed spectrograms for specific plasma parameters (for details, see: Wilson et al. 2008; Elrod et al. 2012; Wilson et al. 2012b, 2013; Nicolaou et al. 2014; 2015a; 2015b). The formula to get the observed counts for an instrument with FOV=($\theta_2$-$\theta_1$)×($\varphi_2$-$\varphi_1$) and energy resolution Δ*E/E*=const is given by Eq.(5) which can be approximated by Eq.(6) for small Δ*E/E*.

In the case of forward modeling is not so trivial to find an analytical expression for the energy at spectrogram's maximum, the behavior of *C(E)* near its maximum and the temperature misestimation when a Maxwell distribution is used instead of a kappa. However, in the Appendix we derive the equations that could be solved to calculate the energy at the spectrogram's maximum in the simplified case where the FOV of the modeled instrument is very narrow and the response of the instrument is not depended on the energy. We show that even in this simplified case, the energy of the spectrogram's maximum is not a function of the plasma temperature only but it depends also on the plasma flow vector. In order to demonstrate the differences of the spectrograms in the case of kappa and Maxwell distribution under the same plasma parameters, we simulate the spectrograms for both distributions and we plot them in the same window. We do this test for two different plasma temperatures and three kappa indices. Our results are shown in **Figure 3**.

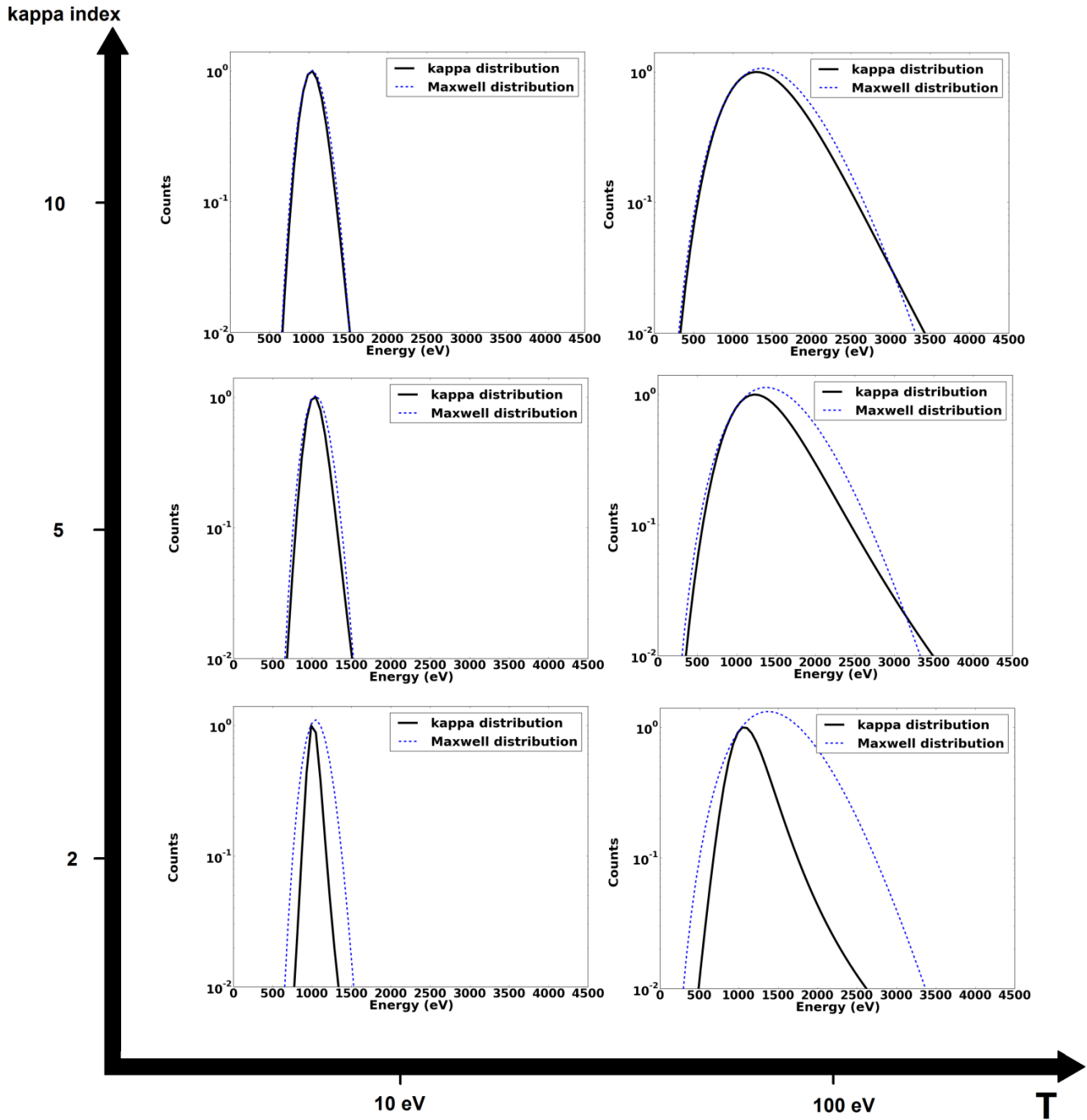

**Figure 3** The counts are depicted as a function of energy for a modeled instrument and plasma described by a kappa distribution (black solid line), and considering Maxwell distribution for the same plasma parameters (blue dashed line). For all the examples the bulk energy is $E_0$=1 keV and the counts are normalized to the maximum counts obtained by the kappa distribution model.

We further want to examine in detail the temperature miss-estimation in cases where the Maxwell distribution is used in the instrument's response model which is used to derive the plasma parameters of plasma that is actually described by a kappa distribution. For this purpose we simulate spin-angle spectrograms of plasma that is assumed to be described by kappa distribution. We simulate spectrograms for several plasma temperatures, several kappa indexes and several hypothetical FOVs of the instrument. We further consider the general case where the instrument is spinning. The spin axis is always considered the symmetry axis of the FOV (see **Figure 4**). In the cases of spinning instruments, we need to study the spin-angle spectrograms. In a spin-angle spectrogram, the x-axis is the spin-phase angle of the instrument, y-axis is the energy, or energy per charge the color represents the counts (McComas et al. 2007; Nicolaou et al. 2014; 2015a; 2015b). We normalize each modeled spectrogram such as its maximum value is 1000 counts for any set of plasma parameters we examine. This is done for more fair treatment among the different spectrograms, since the statistical error of each data-point is the square root of its value. Then those spectrograms are fitted with the forward modeling formula Eq.(6) for a Maxwell distribution in order to derive the plasma parameters. For each fitting, we record the plasma temperature and the reduced chi-squared value which defines the goodness of the fitting. An example of a simulated and a fitted spectrogram is shown in **Figure 5**.

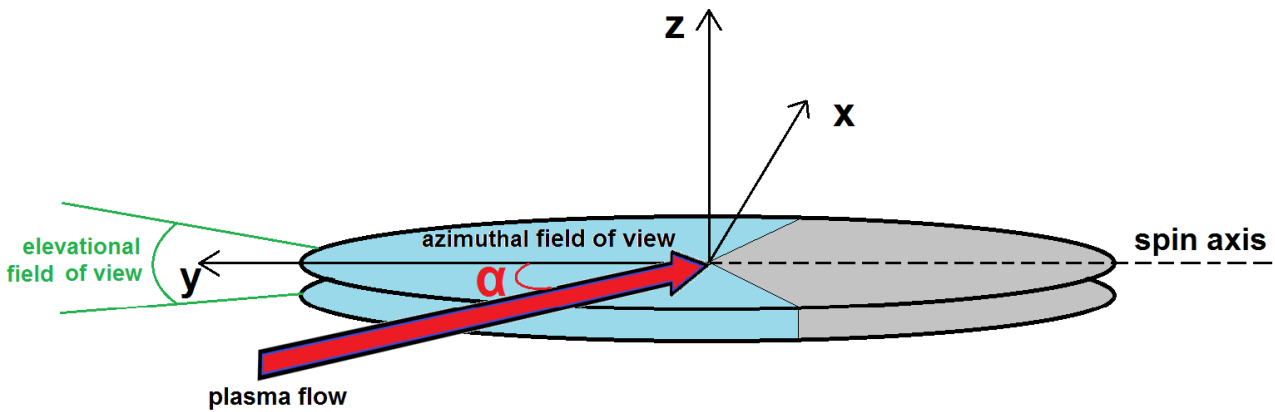

**Figure 4** The hypothetical instrument we use in our study. The light blue shaded area and the green arc represent the angular FOV in azimuth and elevation angle respectively. The FOV is symmetric by the y-axis which is the spin-axis of the instrument. The angle between the plasma flow direction (red arrow) and the spin-axis is denoted by α.

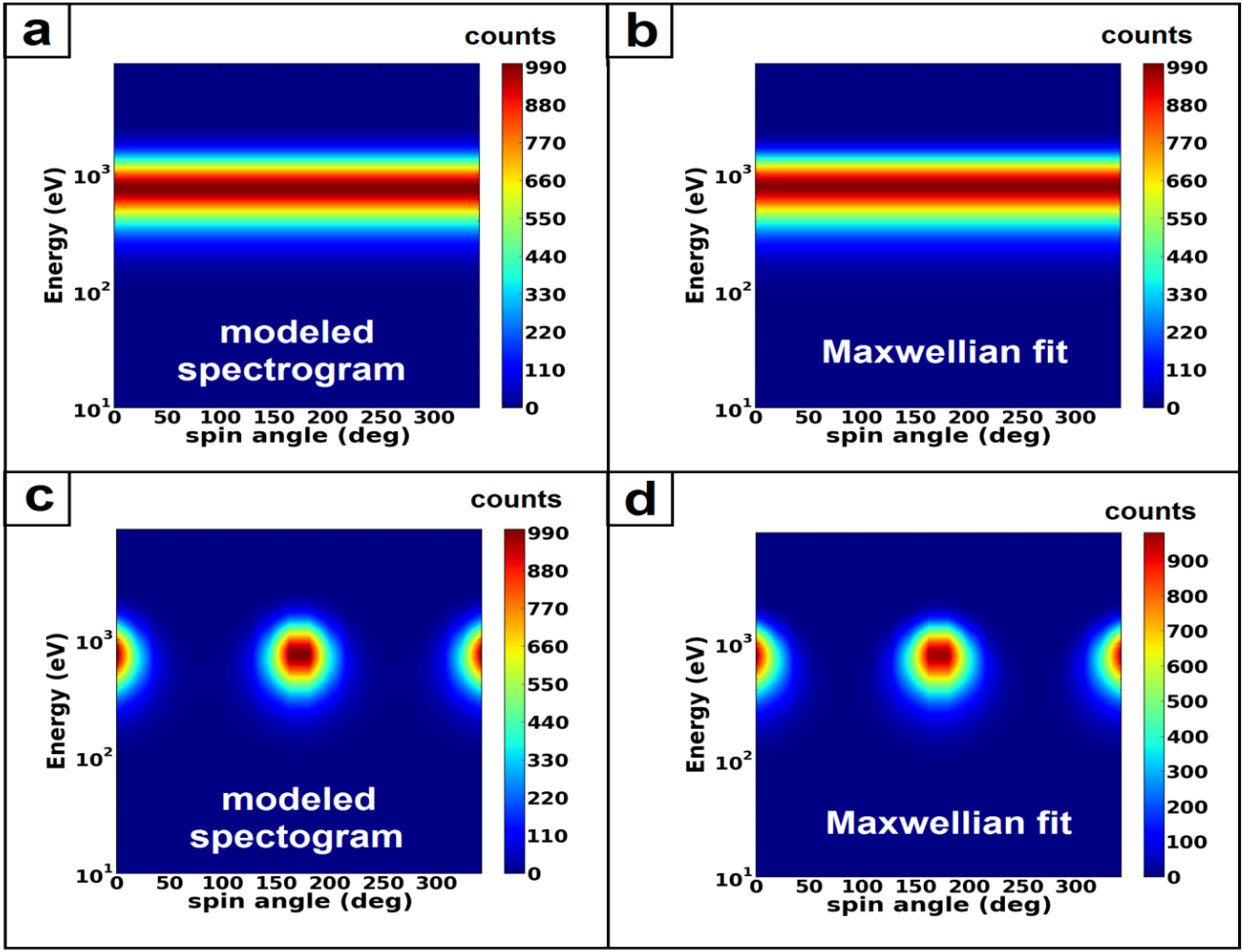

**Figure 5** (a) a modeled spectrogram and (b) a Maxwellian fitted spectrogram of plasma described by a kappa distribution and flow direction parallel to the instrument's spin axis. Panels (c) and (d) show the corresponding spectrograms for the case where the plasma flow vector is 60° apart from the spin axis.

In order to demonstrate the temperature misestimation when the forward modeling method is used to derive the plasma parameters we calculate the ratio of the derived temperature ($T_{M,F}$) over the plasma temperature ($T$) as a function of kappa index, for several $T$ and FOV when the particle beam is directed parallel to the instrument's spin axis. We also examine the reduced chi-square value of each fit which is a straight indication the fit strength. Our results are shown in **Figure 6**. In order to examine any dependence of the results on the beam orientation, we performed the same calculations considering four different beam orientations for one set of FOV and $T$. The corresponding results are shown in **Figure 7**.

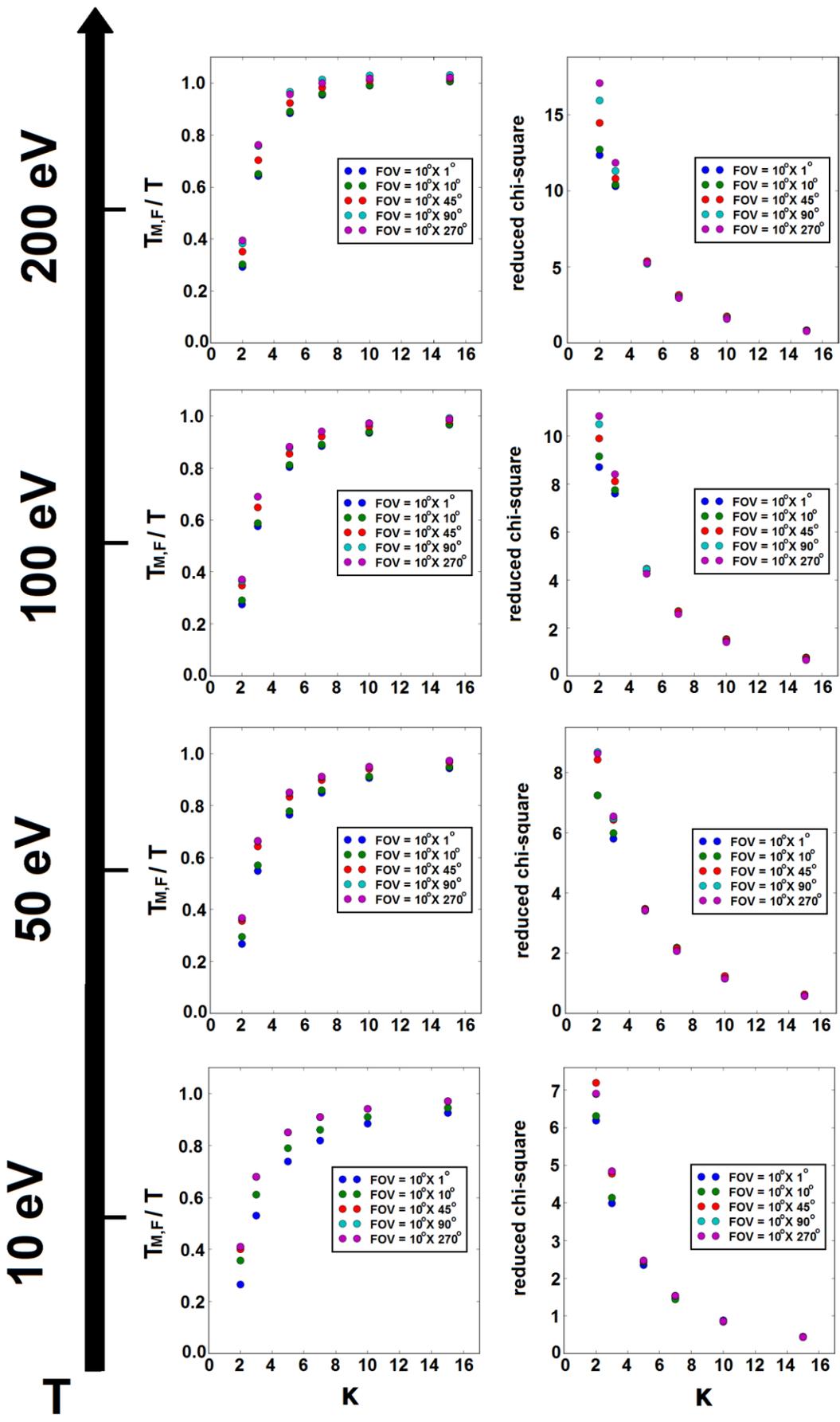

**Figure 6** (Left column) Ratio of the estimated temperature by using the forward model ($T_{M,F}$) over the plasma temperature ($T$), depicted as a function of the kappa index ($\kappa$), and for several plasma temperatures and instrument's FOV. (Right column) The reduced chi-square of the forward model fitting for each panel of the left column.

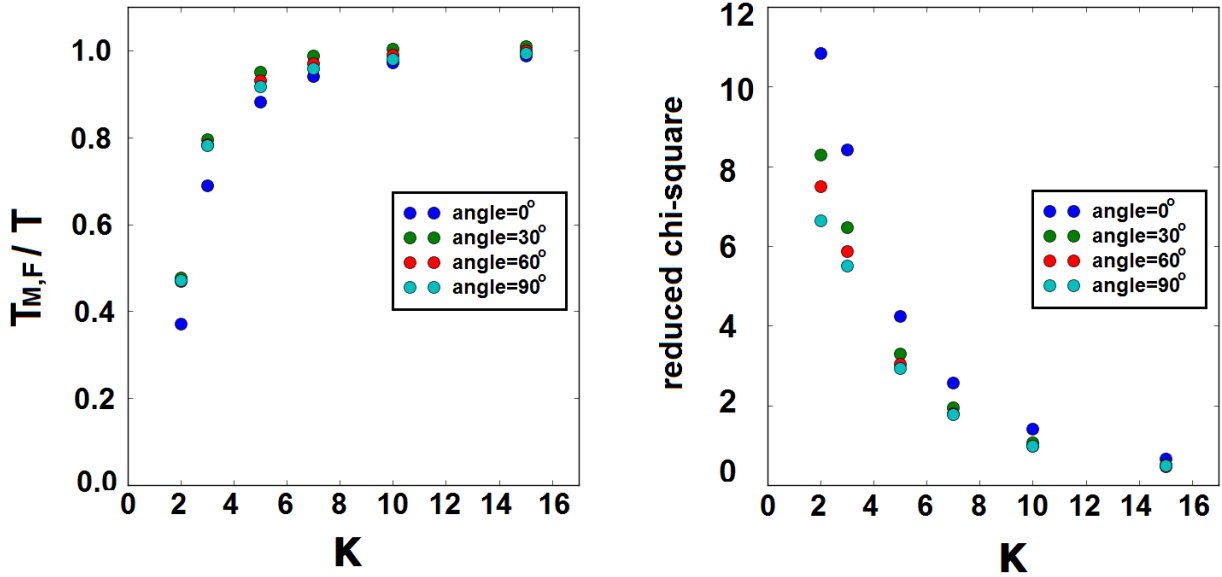

**Figure 7** (Left) Ratio of the estimated temperature by using the forward model ($T_{M,F}$) over the plasma temperature ($T$) as a function of kappa index ($\kappa$), for several directions of the plasma bulk velocity. (Right) Reduced chi-square as a function $\kappa$ for the examined plasma velocity directions.

**4. Discussion and Conclusions**

Diagnostics of laboratory or space plasma systems indicate that the velocity (or the energy) distribution of plasma particles is well described by a kappa distribution function (see references in introduction). This study shows that the plasma temperature can be significantly misestimated when the parameters of a kappa distributed plasma population, of any species, are derived with a Maxwell distribution instead of a kappa distribution function.

This study examined two methods that are widely used for the plasma parameters derivation: (i) the method of direct fitting of the distribution function, and (ii) forward modeling of the instrument's response to simulate energy spectrograms. For the first case, an analytical expression of the plasma temperature misestimation was derived and was verified with several examples. It is concluded that, under this method, the temperature misestimation is a direct function of the kappa index and it is independent of on the plasma temperature itself or the instrument's characteristics. For example, for any plasma temperature we examine, the derived temperature is underestimated by a factor of ~0.17 for $\kappa$~2 and it is underestimated only by a factor of 0.88 for $\kappa$~20 (see **Table 1** and **Figure 2**). When the kappa distribution is fitted with a Maxwellian, the density of the plasma is misestimated as well, leading to the misestimation of other parameters, such as, the plasma thermal pressure. According to (9), the fitted Maxwellian will have normalization constant $C_\kappa = \frac{2}{\sqrt{\pi}} \cdot n(k_B T)^{-\frac{3}{2}} A_\kappa$, which will be misunderstood as the normalization constant of a typical Maxwellian given by (2) with $T=T_M$. Thus, the density will be underestimated by a factor of $A_\kappa \left(\frac{T}{T_M}\right)^{-\frac{3}{2}}$ that corresponds to a factor of ~0.44 for $\kappa$~2 and to a factor of ~0.9 for $\kappa$~20. The thermal pressure for space plasmas, $P = n k_b T$, will also be underestimated (e.g. by a factor of ~0.075 for $\kappa$~2), but interestingly, the polytropic index of the plasma ($\gamma$) will be not. Note that the polytropic index is usually calculated from the slope of $\log(T) = (\gamma - 1)\log(n) + const$. Since both $n$ and $T$ are misestimated by a constant factor (for specific $\kappa$) the slope in the above expression is not changing.

For the second method, the temperature misestimation was demonstrated with several examples of direct fitting modeled spectrograms. Our conclusions are the following:
(i) The misestimation of the plasma temperature, as well as the goodness of the fitting measured by the reduced chi-square, are highly dependent on the kappa index. The temperature misestimation is worst for smaller kappa index. For example, we show cases where the temperature is misestimated by a factor of ~0.3 (for $\kappa$~2) but it is estimated very close to its actual value (for $\kappa$~15).
(ii) For the plasma temperature range examined here, the ratio $T_{M,F}/T$ is closer to 1 for hotter plasma, but the fitting is noticeably worse (higher chi square value for higher temperatures).
(iii) There is a rather strong correlation between the derived temperatures and the instrument's FOV. The dependence is significant for low kappa indices for which the fitting is not good. Interestingly we see that the derived temperatures are closer to their actual values for larger FOV but on the other hand the chi-squares increase.
(iv) The orientation of the particle beam seems to affect the derived temperatures for the cases where the fitting is weak.

We firstly quantified the misestimation of temperature as a function of kappa index (section 3 and appendix). It is shown that the misestimation is larger for smaller kappa index. Note that when the kappa index is high ($\kappa$>10), for all the examples presented, the $T_{M,F}$ over $T$ ratio and the chi-square value are approximately unity and all the data-points of **Figures 6 and 7** pretty much converge to a single-data point.

In hotter plasmas, more particles shift to the higher energies (c.f., Figure 2(b) in Livadiotis and McComas 2010a). The energy and the angular spread of the particle beam increase with increasing temperature. Thus, for hotter plasmas, there are more data-points used in the fitting analysis. Those additional points, due to the distribution's spread, are located further away from the distribution's core (maximum), thus are not well characterized by the Maxwellian form, shown in Eq.(10). This is our explanation why the fitting gets worse for hotter plasma as commented above (see (ii)). The fact that the derived temperatures are slightly higher for hotter plasma is probably due to the fact that we need hotter Maxwellian to fit the points that are spread further from the distribution's core.

As the FOV increases the instrument observes bigger part of the plasma, thus bigger part of the particle distribution which deviates from the Maxwellian form. That is the main reason probably why the chi-square values are higher when the FOV is increased, in the cases of low kappa index ($\kappa<\sim5$) as mentioned in (iii). In the other hand, when the kappa index is high ($\kappa<5$) we notice that the fitting is slightly better for larger FOV. Keep in mind though that for higher kappa index, the Maxwellian spectrogram becomes better approximation for our modeled spectrograms. The strength of the fit may increase when a bigger part of the distribution is taken into account for each data point and when the Maxwellian distribution is a good approximation.

In addition, as mentioned in (iv), the orientation of the beam affects the derived temperatures and the chi-square values when the fitting is weak. When the beam is parallel to the spin axis, then the core of the distribution is inside the FOV for all the spin angles. That means we need to fit more data-points, and when our model uses wrong distribution, the fitting will be weaker.

We remind the reader that in all the cases presented here the chi-square value, which measures the goodness of the fitting, is proportional to the temperature misestimation. Large misestimation of the temperature is always accompanied with large chi-square values. What we really want to carry out from this study is that using the wrong distribution the plasma temperature can be significantly misestimated. In those cases, which are always characterized by weak fitting, the instrument's characteristics and the plasma parameters themselves have their own contribution to the misestimation of the temperature. A modeler should take into account that one of the reasons that are responsible for a weak fitting could be the choice of the wrong distribution. In any case, when there is evidence in the observations, that the plasma distribution deviates from the Maxwellian (high energy tail), a kappa distribution should be tested.

Lastly, we would like to note that the results of this study are for specific plasma conditions, specific instrument model and for normalized spectrograms to a specific maximum value (see section 2). Although the actual values of our results could be different for different instrument or spectrogram normalization, we believe that this study alarms the modelers about the overall dependence of the temperature misestimation on several parameters.

**Appendix A: Energy at the spectrogram's maximum**

In the simplified case, where the instrument's FOV is very narrow and its response is not a function of energy and direction, then Eq.(6) becomes

$$C(E) \propto E^2 f(E;\Omega) . \tag{A1}$$

In order to find the energy at the spectrogram's maximum, we calculate the energy for which the first derivative of (A1) is zero, $\partial C(E)/\partial E = 0$, leading to

$$\frac{\partial \ln f(E;\Omega)}{\partial \ln E} = -2 . \tag{A2}$$

If the observed plasma is described by the kappa distribution, we substitute Eq.(7) and use Eq.(4) in Eq.(A1) that gives the observed counts maximum, which now becomes the quadratic equation

$$(\kappa-1)E_{max} - (\kappa-3)\cos\omega\sqrt{E_0}\sqrt{E_{max}} - 2[(\kappa-\tfrac{3}{2})k_B T - E_0] = 0 . \tag{A3}$$

For small bulk energies, i.e., $E_0 \ll (\kappa-\tfrac{3}{2})k_B T$, $E_0 \ll E_{max}$, the maximum is given by $E_{max} \cong 2k_B T(\kappa-\tfrac{3}{2})/(\kappa-1)$. If we model the instrument's response assuming a Maxwell distribution, we obtain the above again Eq.(A2) but for $\kappa \to \infty$, i.e.,

$$E_{max} - \cos\omega\sqrt{E_0}\sqrt{E_{max}} - 2k_B T = 0 . \tag{A4}$$

and for small bulk energies, the maximum is given by $E_{max} \cong 2k_B T$. Even in this simplified case which we demonstrate here, the location of the spectrogram's peak varies as a function of all the plasma parameters.